\documentclass[sigconf]{acmart}
\usepackage[utf8]{inputenc}
\usepackage[T1]{fontenc}

\usepackage{booktabs} % For formal tables

\usepackage[caption=false,font=footnotesize]{subfig}
\usepackage{fancyhdr}
\usepackage{mdwlist}
\usepackage{multirow}
\usepackage{amsmath}        % enables "cases" for piece-wise functions

\graphicspath{{incl/}}

\makeatletter
\def\blfootnote{\xdef\@thefnmark{}\@footnotetext}
\makeatother

\usepackage{setspace}
\def\smallerspacecaption{\vspace{-2mm}}

\definecolor{gray}{gray}{0.9}

\newcommand{\drop}[1]{}

\renewcommand\footnotetextcopyrightpermission[1]{} % removes footnote with conference information in first column
\makeatletter
\renewcommand\@formatdoi[1]{\ignorespaces}
\makeatother

\begin{document}

\title[An Interposer-Based Root of Trust]{An Interposer-Based Root of Trust: Seize the Opportunity for Secure System-Level Integration of Untrusted Chiplets
}

\author{Mohammed Nabeel,
	Mohammed Ashraf,
	Satwik Patnaik,
	Vassos Soteriou,
	Ozgur Sinanoglu,
	and Johann Knechtel
		}
 \authornote{This work was supported by NYUAD REF (Grant RE218)
		 and by NYU/NYUAD CCS.}
 \affiliation{Tandon School of Engineering, New York University, New York, USA\\
		 Division of Engineering, New York University Abu Dhabi, United Arab Emirates}
 \email{{mtn2, ma199, sp4012, vs86, ozgursin, johann}@nyu.edu}

\begin{abstract}
Leveraging 2.5D interposer technology, we advocate the integration of untrusted commodity components/chiplets 
with physically separate, entrusted logic components.
Such organization provides a modern \emph{root of trust} 
for secure system-level integration.
We showcase our scheme by utilizing industrial \emph{ARM} components
that are interconnected via a security-providing active interposer,
     and thoroughly evaluate the achievable security via different threat scenarios. 
Finally, we provide detailed end-to-end physical 
design results to demonstrate the efficacy of 
our proposed methodology.
\end{abstract}

\maketitle

\vspace{-2mm}
\section{Introduction}
\label{sec:introduction}

Attackers have traditionally focused
on software and their vulnerabilities (and will also do so in future). Nowadays, tackling the underlying hardware
is also becoming promising for an adversary, e.g., due to ``bad decisions'' made by designers many years ago~\cite{lipp18}.
Besides various works addressing these risks at design- and manufacture-time, there are many hardware-based countermeasures seeking to mitigate
security threats at runtime.
Among many, they include enclaves for trusted execution, e.g., \emph{ARM TrustZone} or \emph{Intel SGX},
wrappers for monitoring or cross-checking of third-party intellectual property (IP) modules~\cite{basak17,chandrasekharan15},
centralized security infrastructures~\cite{wang15_IIPS},
secure task scheduling~\cite{liu15_TETC}, and secure architectures for Networks-on-Chips
(NoCs)~\cite{fiorin08,caimi17,wassel14}.

Considering the ongoing trend for outsourced manufacturing in the chip industry,
there are many steps in the supply chain in which an adversary can penetrate so as to compromise the security of electronic circuits~\cite{rostami14}. Hardware Trojans~\cite{xiao16} or, more broadly, any malicious modification introduced during IC fabrication represent serious hazards. Hardware security features and components are arguably the most vulnerable targets here as adversaries choose to bypass or disable them, and, to the best of our knowledge, no prior art has demonstrated that they can withstand malicious modifications. Besides, before deployment of security-feature-enriched chips (or any chips), their verification against 
modifications requires sophisticated solutions (e.g.,~\cite{subramanyan14}),
not least because physical verification is difficult in general~\cite{wang17_data}.
In other words, any hardware security feature itself becomes prone to being stealthily misused, or circumvented altogether, once the
adversaries have access to the supply chain, particularly during IC
manufacturing.\footnote{Remarkably, this concern may also apply to
advanced packaging or 3D integration. For example, Valamehr {\em et al.}~\cite{valamehr10,valamehr13} propose a security monitor to be 3D-stacked on top of a commodity processor. Their monitor is based on security features such as tapping and re-routing, which all rely on introspective interfaces within the commodity processor and its components.  These
features may fail or be mislead with false data in case those interfaces are ``hacked'' by malicious third parties involved for the design and/or manufacturing of the commodity chip~\cite{
	valamehr13}.

}

Irrespective of security-compromising issues in ICs, advanced packaging and 3D integration technologies have made significant progress over
recent times~\cite{clermidy16,pavlidis17}.
This umbrella of diverse technologies collectively embrace the notion of ``building city clusters or skyscrapers of electronics.'' Silicon \textit{interposer technology}, otherwise known as ``2.5D stacking,'' facilitates chip-level integration by placing die side-by-side; it therefore serves as an integration carrier and accommodates a specific underlying system-level interconnect fabric to provide inter-die communication~\cite{lau11,pavlidis17,clermidy16}, resembling a modern version of a printed circuit board. Building an advanced electronic system using an interposer is typically less costly and less complex than native 3D
integration~\cite{stow17,pavlidis17,lau11}.
In fact, interposer-based systems are already prominent in the market~\cite{lee16,shilov18}.

\begin{figure}[tb]
\centering
\includegraphics[width=.85\columnwidth]{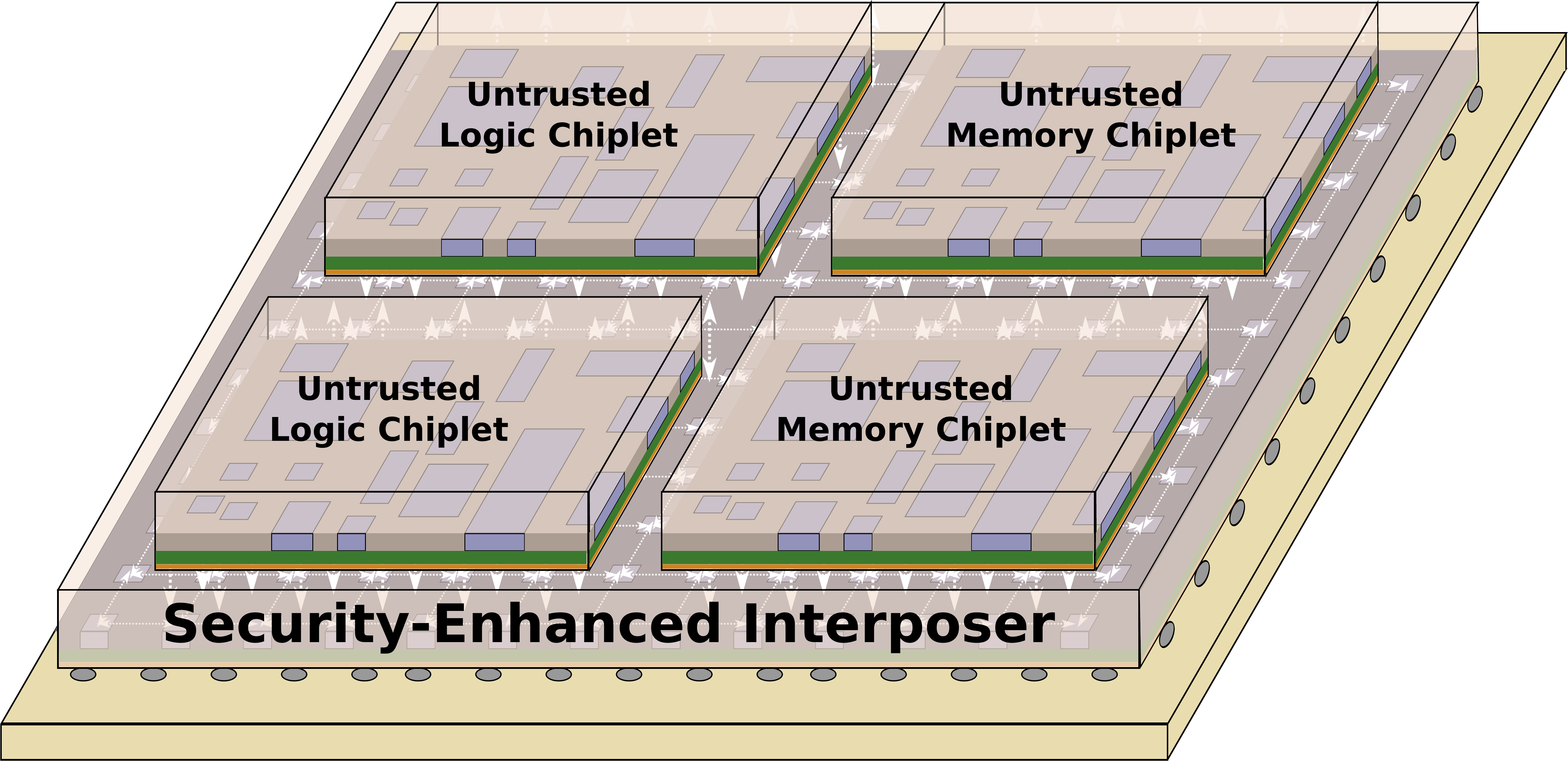}
\smallerspacecaption
\caption{
An interposer acting as the security backbone or ``root of trust'' for system-level integration of untrusted chiplets.
Any system-level and external communication of the chiplets is physically enforced to be routed through the root of trust, where every
transaction is monitored and either accepted/passed through or denied/blocked.
\label{fig:concept}
}
\smallerspacecaption
\smallerspacecaption
\smallerspacecaption
\end{figure}

Concurrently, there are strong efforts to drive the concept of design reuse, from IP modules 
to reusing whole chips at the system level.
In particular, the \textit{DARPA CHIPS} initiative~\cite{CHIPS}
targets for
common interface standards and system-level integration of \textit{chiplets}, i.e., chips encapsulating some, more or less, complex functionality
(e.g., microprocessors) and implementing such standard interfaces.
The notion of chiplet integration has been well-received by the academia (e.g.,~\cite{yin18,stow17,coskun18}) and the industry;
related products are already in the market, e.g., the
\textit{AMD Epyc} chip~\cite{shilov18} or the \textit{Intel Embedded Multi-Die Interconnect Bridge} technology~\cite{mahajan16}.
Naturally, advanced packaging and 3D integration technologies are promising further advancement of such system-level integration of chiplets.

Now, we believe that seizing the opportunities provided by advanced packaging and 3D integration is timely to, quite literally,
open up a new dimension for the design of secure and
trustworthy electronic systems.
This paper can be summarized as follows.
\begin{itemize}
\item Drawing help from prominent interposer technology, we propose a novel concept for a modern ``root of trust,''
offering a clear physical separation between commodity and security components (Sec.~\ref{sec:concept}).
We integrate untrusted commodity chiplets and separate security components (monitoring the chiplets at runtime),
along with a system-level interconnect fabric. Towards this end, we make use of an interposer as a physically separate, trustworthy backbone.
\item Our concept does not require any trust assurance for the commodity chiplets---the chiplets may run malicious code and/or
contain some hardware Trojan(s), all without undermining the system-level trustworthiness of our scheme.
\item We showcase our concept using industrial components from \emph{ARM}, namely the \emph{Cortex-M0} chiplet and the \emph{Advanced
Microcontroller Bus Architecture High-Performance Bus (AHB)}, to implement a secure multi-core system with distributed shared memories
(Sec.~\ref{sec:implementation}).
Here we also develop dedicated security features for memory access and data control.
\item We thoroughly evaluate our case study (Sec.~\ref{sec:experiments}). We implement different security policies and
demonstrate them in action against malicious runtime behavior (at the simulation level).
We also conduct a full, end-to-end physical design of our case study, and we
elaborate on the acceptable layout cost of our proposed security scheme.
\end{itemize}

\section{Concept and Threat Model}
\label{sec:concept}

Among the various options for advanced packaging and 3D integration, here we pursue the use of the 2.5D interposer technology.
More specifically, we leverage an interposer as \textit{security backbone} (Fig.~\ref{fig:concept}).
Such a \textbf{concept} offers some important benefits as follows:

\begin{enumerate}
\item
An interposer enables ``plug and play integration'' of untrusted components/chiplets and security modules
with a \textit{clear physical separation} between those components.
As of now, using an interposer is the \textit{sole option for secure system-level integration} of
multiple components. Prior schemes focused on classical 2D integration that cannot tolerate malicious modifications of their security features.

\item
An interposer acting as a security fabric allows us to impose the policy that \textit{any untrusted component has to depend on this backbone} for functionality provision and/or system-level communication, and not vice versa.
Hence, we do not require any chiplets to provide security assurance, and we also avoid risking any interference with the security features.
Once adversarial activities are observed by security monitors, the related communication is
blocked directly at the backbone.
Thereby, the system experiences ``only'' a loss of functionality, but its \textit{integrity and trustworthiness remain intact}.

\item
An interesting option is to implement an \textit{active interposer} where the
integration carrier also contains some logic, not only passive wiring.\footnote{Active interposers have been successfully demonstrated in the past~\cite{stow17,clermidy16,takaya13,lau11}. One can simply think of them as regular chips, with relatively large outlines, but also of relatively simple logic and low utilization rates (thereby managing yield).  Therefore, they may also be implemented using an older technology node. In our work, alternatively to using an active interposer, the security features can also be implemented in another chiplet, leveraging a trusted facility for its fabrication. Then, the system-level interconnect fabric in the passive interposer has to be designed such that the security chiplet serves as central routing node. This may
give rise to bandwidth limitations; optimizing the interposer interconnect is an active area of research
itself~\cite{
	akgun16,
	kannan15,
	yin18}.}
This way, the \textit{security features can be implemented into the
interposer itself}, which allows for dedicated and direct
	monitoring of all individual chiplets.

\end{enumerate}

Naturally, the fabrication of the interposer has to be trusted---any risk of malicious modifications cannot be tolerated.  Given that the
interposer comprises only the system-level interconnect and possibly the security features (i.e., when using an active interposer, as we propose in this work), the design and manufacturing costs are considered manageable and reasonable.

\textbf{Threat model:} There are various threats to be considered when seeking to securely integrate components at the system
level~\cite{basak17}.  In general, all threats concern the system-level communication and system-level behavior.  More specifically, a
malicious component may act as (i)~passive reader (\textit{snooping}), (ii)~masquerader (\textit{spoofing}),
(iii)~modifier, and/or (iv)~diverter or re-router~\cite{basak17}.

Our proposed concept is the first (to our best knowledge) that can rule out all above threats by construction. As for points (i), (iii), and (iv), their validity applies as the system-level interconnect fabric is physically separated from all untrusted components---these components cannot access any communication not directly addressed to or created by them. As for point (ii), we delegate the interconnect interfaces, which also handle memory addresses, completely to the security backbone 
(Sec.~\ref{sec:implementation}). In this way, the essence of spoofing, that is, the malicious modification of source or destination addresses, cannot be achieved by 
untrusted components to begin with.

As we implement a multi-chiplet system with shared memories (Sec.~\ref{sec:implementation}),
we still have to address the generic threats of malicious accesses and modifications of memory-resident data carried out at runtime. Here, we assume that such adversarial behavior is either introduced by untrusted chiplets, or by some malicious software, or by some
hardware Trojan(s). Any attack is considered to be exercised through system-level communication targeting the shared memory.
That is, as of now, we do not account for side-channel based attacks or any Trojans within the
memory itself.\footnote{As for side-channel
attacks
prior art by Indrusiak \textit{et al.\ }\cite{indrusiak16} suggests randomization of routing. Such
	techniques can be readily included in our scheme as well.
}

\section{Architecture and Implementation}
\label{sec:implementation}

Here we describe the functionality and physical design of our proposed \textit{interposer-based security-enhanced architecture}, dubbed
\textit{ISEA} for short.\footnote{More colloquially, we can also think of and memorize ISEA as ``I see (y)a'' which reflects upon the system-level security monitoring enabled by ISEA.}
The interposer acts as the ``root of trust,'' i.e., it checks the legitimacy of every requesting communication transaction and accordingly enforces security policies upon them to protect the entire system in case of detected malicious activity. Key to our design is the \textit{transaction monitor (TRANSMON)} which administers said security policies, comprising an \textit{address protection unit (APU)} and a \textit{data protection unit (DPU)}. Both the APU and the DPU work in tandem to collectively serve the establishment of our security-enhancing policies, as detailed next.
We plan to release our proof-of-concept implementation to the community.

\subsection{ISEA-Based System Architecture}\label{subsec:SysArchDescription}

We describe the implementation of ISEA in the context of the widely-used \textit{ARM AMBA} on-chip bus architecture. We mote that we exhibit only a particular instance of ISEA here, as such a security-enhanced interposer can be retro-fitted to secure other existing (or future) multi-chiplet architectures as well. We focus on the \textit{AMBA Advanced High Performance Bus (AHB)} that serves components which demand high communication throughput, such as processors and memories.
AHB enables communication among bus-attached master
components that initiate bus data transfers and bus-attached slave components that respond to the requests of the masters. The AHB bus transfers data values, addresses, and control info, managed by logic components such as arbiters, decoders, multiplexors, etc., all of which basically implement the AMBA protocol.

\begin{figure}[tb]
\centering
	\includegraphics[width=\columnwidth, trim = {0mm 10cm 11cm 0mm}, clip=true]{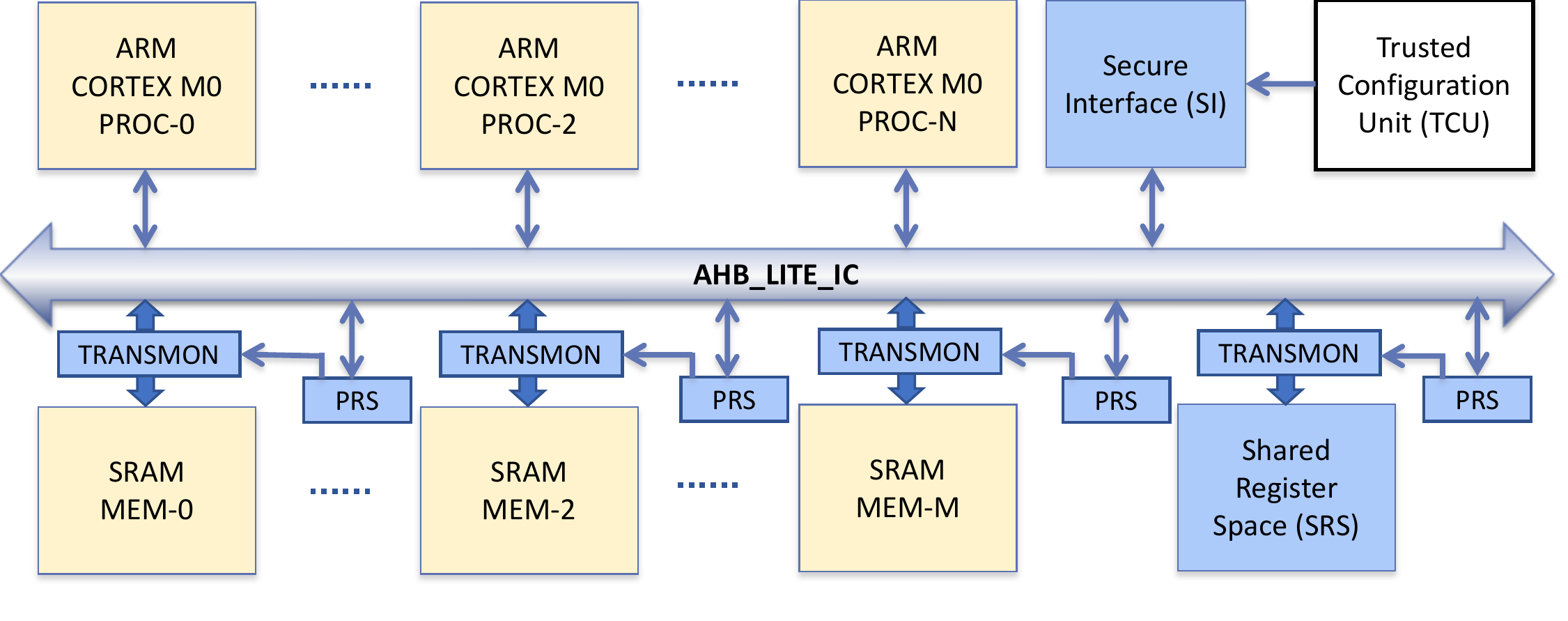}
\smallerspacecaption
\smallerspacecaption
\smallerspacecaption
\smallerspacecaption
\caption{Overview of ISEA.
	Blue parts are implemented exclusively in the
    ``root of trust.''
		A policy register space (PRS) holds the security policies relevant for each transaction monitor (TRANSMON). Only the trusted configuration unit (TCU), via the secure interface (SI), can update the PRS.
\label{fig:architecture}
\smallerspacecaption
\smallerspacecaption
\smallerspacecaption
\smallerspacecaption
}
\end{figure}

Figure~\ref{fig:architecture} depicts the proposed ISEA architecture. It should be noted that the AHB and the various proposed security-enhancing constituent components such as the SRS, the SI, and all TRANSMONs along with their PRSs, are all implemented exclusively in our security-enforcing
interposer; the functionality and physical implementation details of all these modules will be described shortly in Section~\ref{subsec:PhysicalDesign}. Without loss of generality, here we utilize AHB to interconnect four core chiplets (each comprising 16 ARM M0 cores), along with four SRAM memory chiplets, nevertheless, as the architecture is scalable, any other reasonable number of chiplets and cores/chiplet can be incorporated into the system to meet the desired computational demands. All core and memory modules, which constitute off-the-shelf plug-in components, are interconnected to a shared-memory multi-chiplet environment to support distributed computation where cores are allocated during runtime as defined by the OS that dictates system-level thread/process scheduling. 

All SRAM modules (i.e., slaves) occupy the entire memory space, with each being distributed a subset of the memory map, where communication transactions are implemented as reads or writes to specific memory locations. Chiplets (i.e., masters) request access to the bus in order to load/store data from/to the shared memory.

\subsubsection{Transaction Monitor (TRANSMON)}\label{TRANSMON_Describe}

\begin{figure}[tb]
\centering
	\includegraphics[width=.95\columnwidth, trim = {0cm 0cm 2cm 0cm}, clip=true]{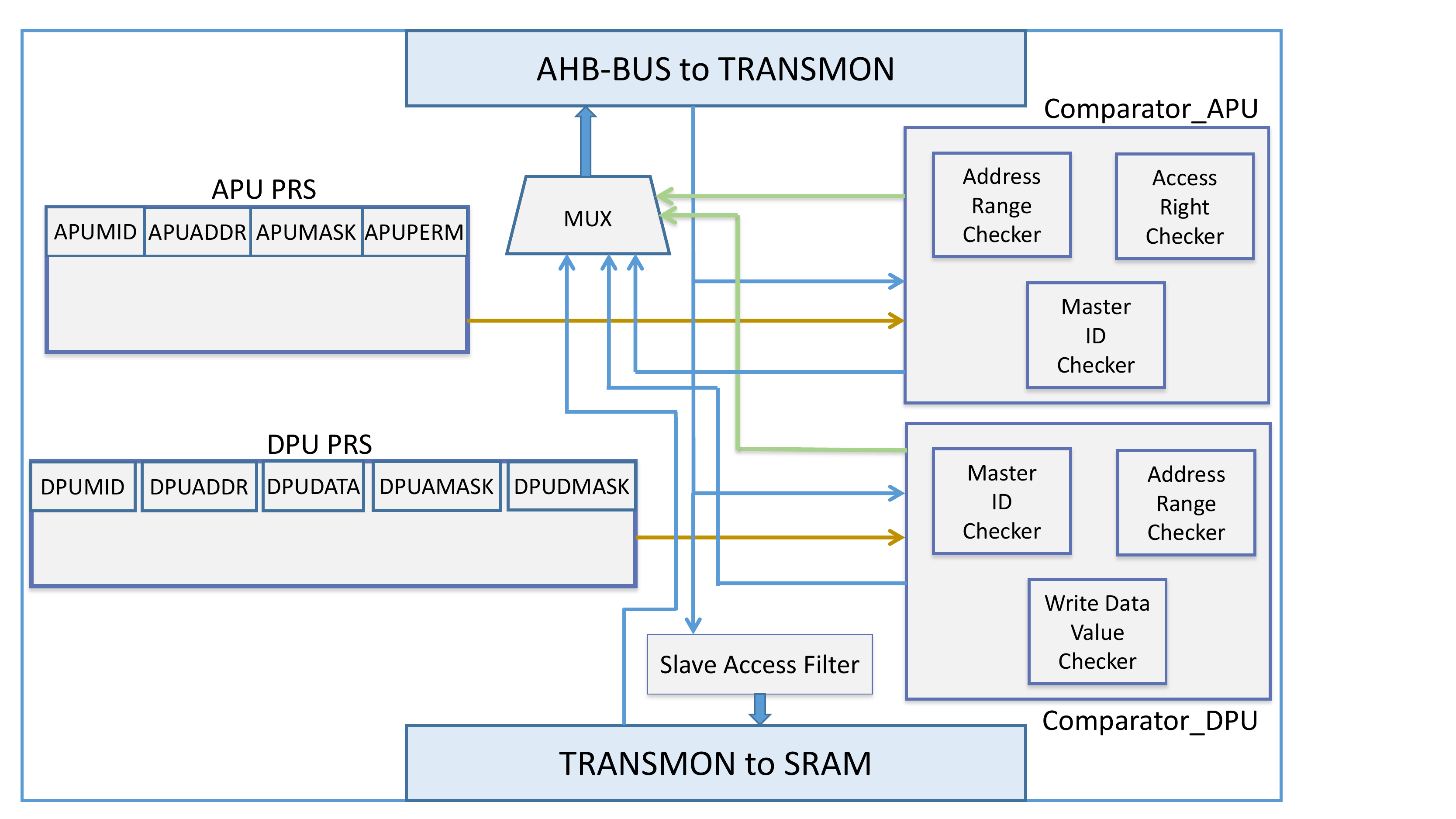}
\smallerspacecaption
\smallerspacecaption
\caption{Zoom-in of the TRANSMON features.
\label{fig:TRANSMON}
}
\smallerspacecaption
\smallerspacecaption
\smallerspacecaption
\end{figure}

Key to our design is the TRANSMON (Fig.~\ref{fig:TRANSMON}), with one placed in-between every slave component and the AHB
bus interface.\footnote{Another option would be to place the TRANSMONs between the master components and the AHB bus. However, our design
	decision offers two important benefits. First, a TRANSMON at the master would require additional address decoding, which is already
		covered by AHB itself. Second, a TRANSMON at the slave allows to keep track of only the policies relevant to that slave. As
		a result, for our architecture, the hardware cost and delays imposed by policy checking are restricted.} This serves to continuously check upon the legitimacy of every transaction by the TRANSMON attached to the specific SRAM that contains the memory addresses upon which read or write accesses are requested, after such transactions successfully complete AHB arbitration and routing.

A TRANSMON blocks transaction requests that either (a) do not possess the permissions to access shared memory regions, as dictated by the APU, or (b), are not allowed to write out specific data, as determined by the DPU; by default, TRANSMON also blocks any request that cannot be matched to some policy. Also note that all policies involve the handling of master/slave IDs. Since the TRANSMON-AHB interfaces are all implemented in the
interposer, there cannot be any spoofing of IDs to begin with.
Hence, the APU protects against any undefined and/or malicious access while the DPU protects against any illegal data modification or leakage.
In case a request is denied, the TRANSMON passes an error message to the chiplet (master) which initiated the transaction, while the
memory access is simply dropped (by the \textit{Slave Access Filter}), thereby protecting the data.

A TRANSMON comprises an APU, a DPU, and some glue logic.
Both the APU and DPU have their own \textit{policy register space (PRS)}.
For efficiency, the PRSs are implemented using flip-flops.
Each PRS entry defines an individual policy concerning either
(a)~some particular region in the shared memory space, physically allocated in the related memory slave connected to that TRANSMON,
or (b)~some particular data.
The PRSs are initialized from the \textit{trusted configuration unit (TCU)} which has privileged access to the AHB via the \textit{secure interface
(SI)}.
The TCU is required for other system-level steps,
    namely the scheduling of the multi-core application, the loading of the initial data into the shared
memories, the compilation of all required policies, as well as sending the interrupt/request to start the computation.
Note that the TCU can be provided either by an external, trusted entity or it can also be realized by a fully trusted processor directly
stacked onto the interposer.

\subsubsection{Address Protection Unit (APU)}\label{subsec:DPUinfo}

The APU forms an essential part of the TRANSMON and for the proposed ISEA architecture.
As such, it is also to be implemented in the interposer ``root of trust.''
Each APU makes use of its own PRS, which is also embedded in the interposer.
The function of the APU is to control any chiplet requests against specific memory ranges.
The control covers the type of access
(i.e., whether it is read-only, write-only, or read/write) and the actual address or address range.
A simple 2-bit encoding is used for read/write access, with ``01'' for read-only (RO), ``10'' for write-only (WO)
and ``11'' for read/write (RW) (``00'' is reserved).
For address ranges, address masking is leveraged.

As Fig.~\ref{fig:TRANSMON}
shows, an APU policy comprises four parameters: (1) Master ID (APUMID), which identifies the master chiplet
which initiates a memory request;
(2) a 32-bit address (APUADDR);
(3) a 32-bit address mask (APUMASK), where the transaction is valid only if HADDR if greater than APUADDR
logically bit-wise ANDed with APUMASK, where HADDR is calculated as the logical ORing of APUADDR with APUMASK;
(4) the read/write permission (APUPERM), i.e., whether RO, WO, or RW.

\subsubsection{Data Protection Unit (DPU)}\label{subsec:DPUdescribe}

The DPU, like the APU, forms an essential part of the TRANSMON.
Its function is to provide data-level protection in the multi-chiplet ISEA system.
This is achieved by blocking (a) overwriting of data in the event of unauthorized writes to specific memory locations, or (b) writing out
particular data.  The former serves protection of data at runtime, and the latter serves protection of \textit{assets}, e.g., secret keys,
from leaking into the shared memory.
Superficially, a transaction is not allowed to proceed when the DPU PRS
contains a policy that disables writing of particular data to some address range.
Note that the DPU incurs one cycle delay in all the write transaction to the slave, as the DPU checks can only
work during the data phase of the AHB protocol.
Hence, until the data is checked, address and control signals are registered and kept.

As Fig.~\ref{fig:TRANSMON} shows, a DPU policy comprises five parameters:
(1) Master ID (DPUMID);
(2) a 32-bit DPU address (DPUADDR);
(3) a 32-bit restricted-write data value (DPUDATA);
(4) a 32-bit data mask (DPUDMASK), which serves to block the transaction in the event
that the logical ANDING of HWDATA with the bit-wise inversion of DPUDMASK equals to DPUDATA;
(5) a 32-bit address mask (DPUAMASK), with the working mechanism as in the APU.
        
\subsection{Physical Design} \label{subsec:PhysicalDesign}

Here, we elaborate on our end-to-end flow of our proposed interposer-based multi-chiplet design. 
The physical design flow encompasses two major steps, next discussed in detail. 

\begin{figure*}[tb]
\subfloat[Floorplan of ISEA. 4 ARM chiplets in the middle, 4 memory cores around.
Inset: Microbumps in yellow for the interposer.]{
	\includegraphics[height=60mm]{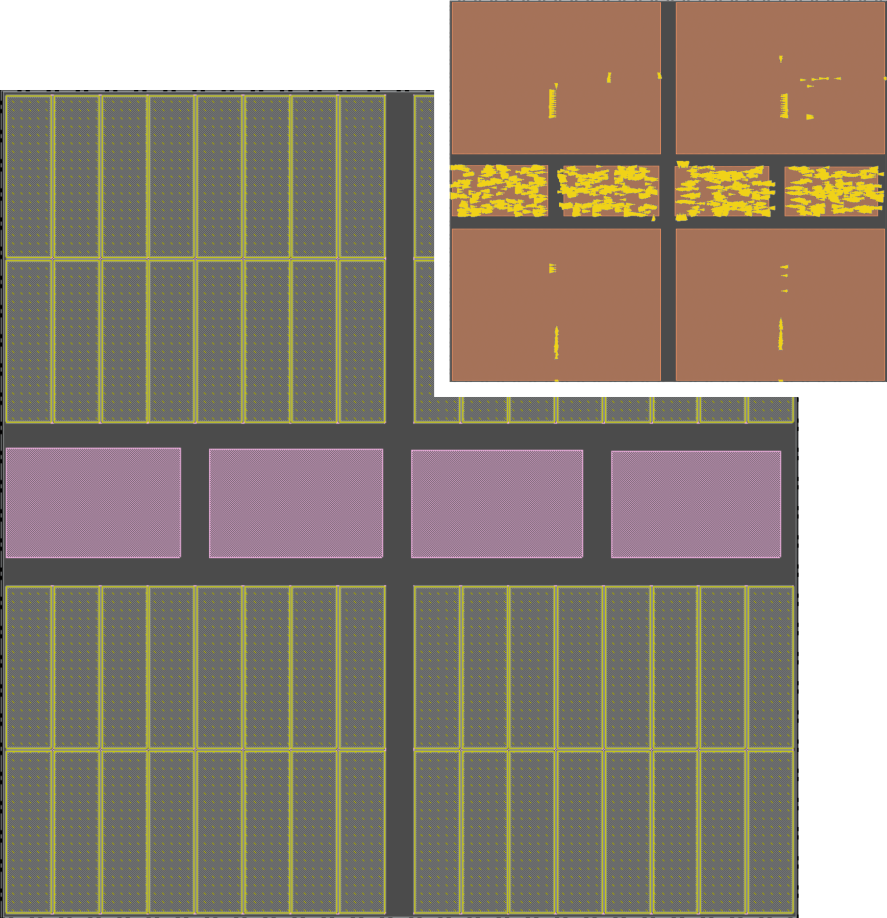}
}
\hfill
\subfloat[ARM M0 chiplet (16 cores). Logic and microbump locations (top), routing (bottom).]{
	\includegraphics[height=66mm]{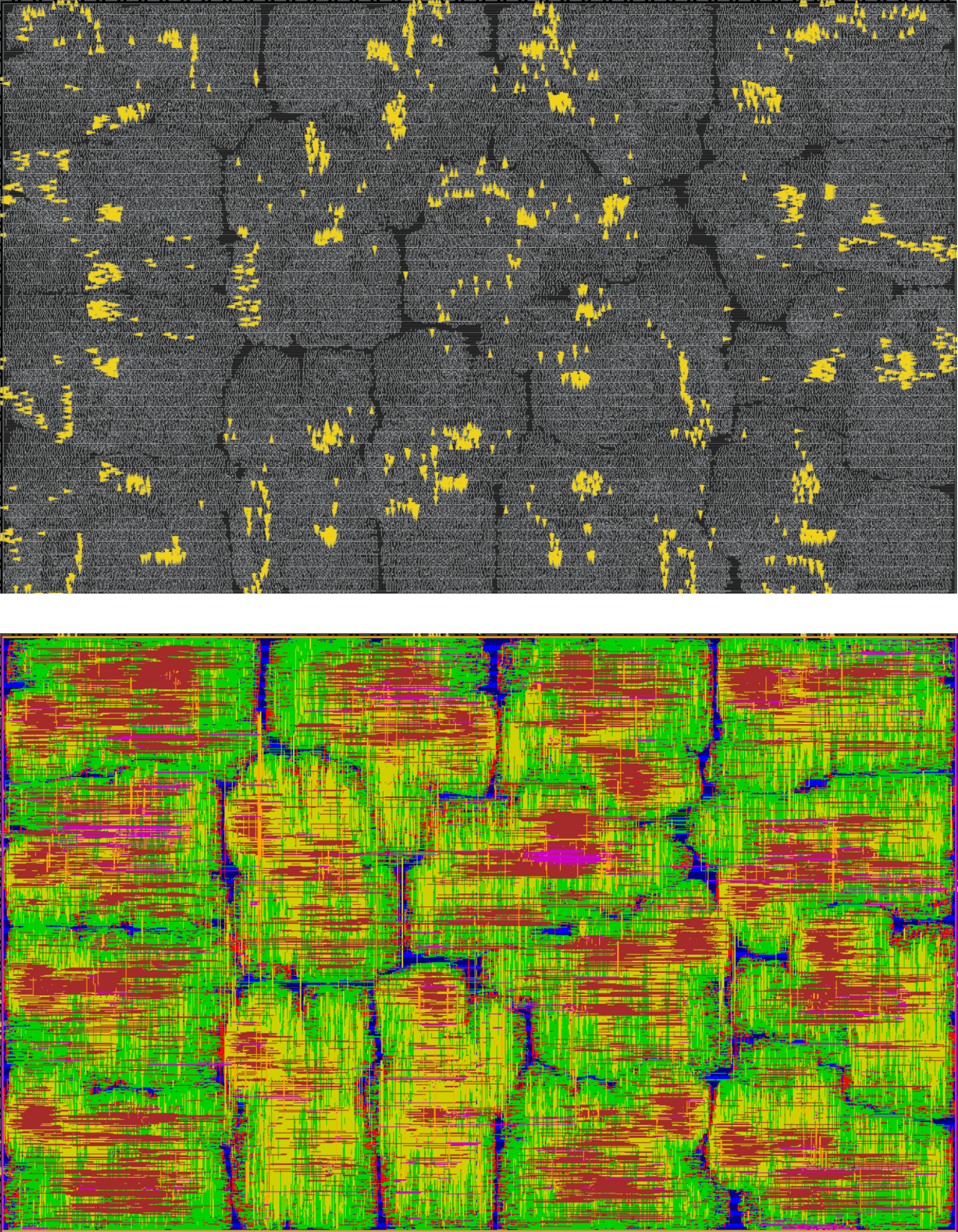}
}
\hfill
\subfloat[ISEA implementation in interposer.
TRANSMON in white, regular AHB components in grey.
Inset: Routing across four metal layers.
]{
	\includegraphics[height=66mm]{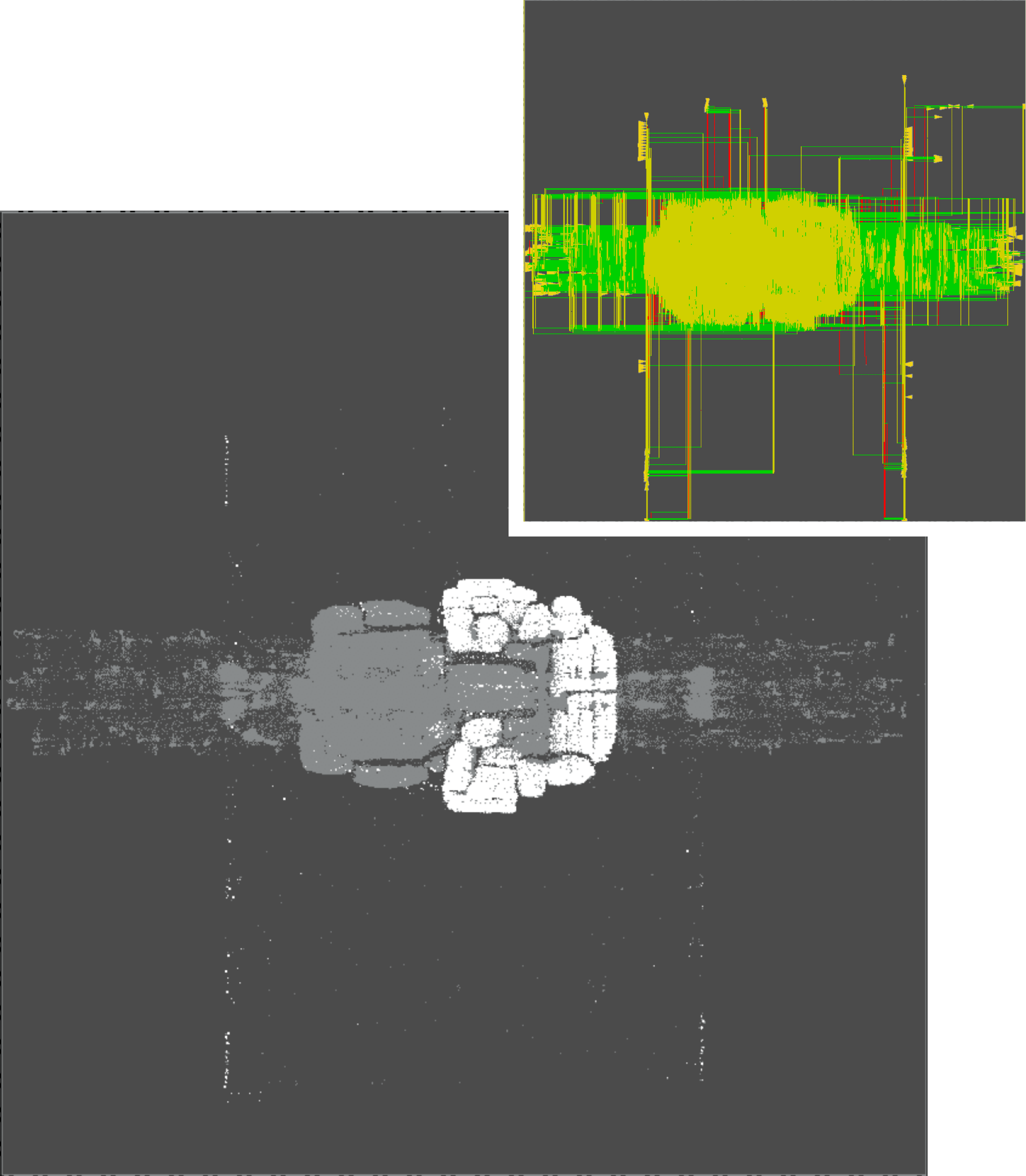}
}
\smallerspacecaption
\caption{Layout snapshots, obtained from \textit{Cadence Innovus}.
\label{fig:layouts}
}
\smallerspacecaption
\end{figure*}

\textbf{Design Partitioning and Floorplanning:}
We partition ISEA's full chip netlist into multiple chiplets of core/memory banks, where a core (memory) bank contains multiple processor
cores (memories), the size of which are set by the user. As indicated earlier, for our proof-of-concept design we implement 4 core
(master) banks, each comprising 16 \emph{ARM-Cortex M0} cores, and 4 memory (slave) banks, each containing 16 64KB memories.
In our flow, the designer has the freedom to specify the size of the core and memory banks, the starting
utilization, and the aspect ratio individually, which is all passed to the \emph{floorplanning} procedure. 
Essentially, the floorplanning matrix appears in
the following order: \{\{MB MB\} \{CB CB CB CB\} \{MB MB\}\},
where MB stands for memory bank and CB for core bank;
Figure~\ref{fig:layouts} shows this arrangement.
The floorplanning also generates placement information for the memory blocks in the MBs which are used for the 2D memory implementation
based upon designer inputs concerning the horizontal and vertical spacing between memories.
The floorplan information is saved in \emph{TCL} format and then utilized in the 2D implementation of the chiplets, including the interposer.
Chiplet netlists are then generated, timing budgets are derived, and, finally, chiplet constraints are calculated.

We place emphasis on the fact that the physical design flow is flexible concerning the chiplets being designed in-house or procured as
hard IP. For hard-IP chiplets, the steps are more straightforward, and essentially covering the interposer design.
That is,
the designer has no freedom in modifying the size of the chiplet.

\textbf{Planning of Interposer Microbumps and Design Closure:}
If not procured as hard IP, individual chiplets
proceed through the standard 2D implementation flow which generates individually routed layouts
with microbump locations. 
For the interposer microbumps, their locations are chosen around the vicinity of drivers/sinks.
On-track \textit{legalization} is performed for the interposer microbumps via custom scripts; the objective here is to maximize the
utilization of routing resources.
RC parasitics are generated from the \emph{postRouted}, final layout in SPEF
format. This file is then used for sign-off analysis, i.e., to evaluate power and performance. 

Once the 2D implementation for all chiplets is done, or immediately in case the chiplets are hard IP,
the interposer P\&R flows follows.
First, the interposer netlist---essentially containing ISEA---is imported along with the microbump locations derived from all chiplets.
After completing the regular design flow of the interposer and exporting the final netlist,
we extract the final RC parasitics and stream out the respective GDS.
To evaluate ISEA's chip-wide PPA, a wrapper netlist is generated for
   the interposer and all the chiplets. 
   The interposer microbump parasitics are modelled into the SPEF of the wrapper.
   We do not engage in any inter-tier cross-optimization; again, that is essential to accommodate for hard-IP chiplets.

\section{Experimental Evaluation}
\label{sec:experiments}

\textbf{Experimental setup}: We implement ISEA end-to-end for a 64-core multi-chip system using \emph{ARM} M0 cores, showcasing the practicality of our work. 
The 64 cores are organized into four chiplets, each holding 16 cores; we also employ four memory chiplets with 1 \textit{MB} SRAM for each. 
We emphasize that this implementation constitutes a proof-of-concept; in reality, larger memory chiplets and more complex cores are likely to be utilized. 
The Register Transfer Level (RTL) code for the complete system including the TM, bus matrix etc., have been implemented using \emph{Verilog}. 
Verification has been performed using \emph{Synopsys VCS}, while \emph{ARM's} IAR workbench is used to generate the processor program codes. 
Synthesis was performed using \emph{Synopsys DC} and layout generation using \emph{Cadence Innovus v.17.10}.
We leverage the 65 \textit{nm} \emph{Global Foundries} technology 
and ARM standard cell and memory libraries.
We also utilize the \emph{Synopsys} SAED 90 \textit{nm} technology to synthesize a second version of the interposer.
We utilize 7 and 4 metal layers for core and interposer respectively, for both 65 and 90 \textit{nm} technologies.

\subsection{Security Evaluation}

\begin{figure*}[tb]
\subfloat[Malicious read access blocked by an APU policy.]{
	\includegraphics[height=5cm, trim = {0mm 0cm 14cm 0mm}, clip=true]{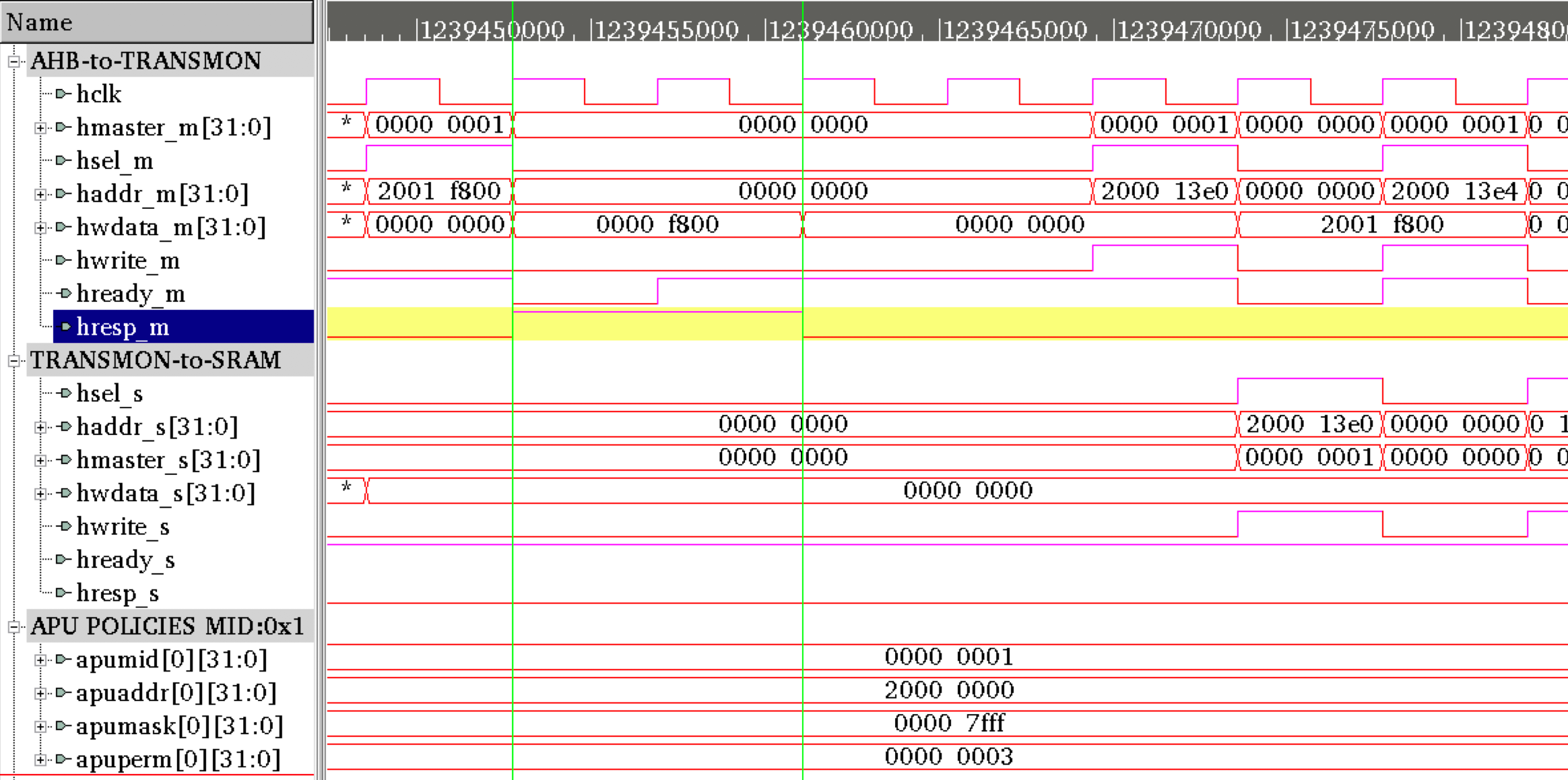}
}
\hfill
\subfloat[Malicious write-out of a secret key \textit{0XBAD\_BEEF} blocked by a DPU policy.]{
	\includegraphics[height=5cm, trim = {0mm 0cm 0cm 0mm}, clip=true]{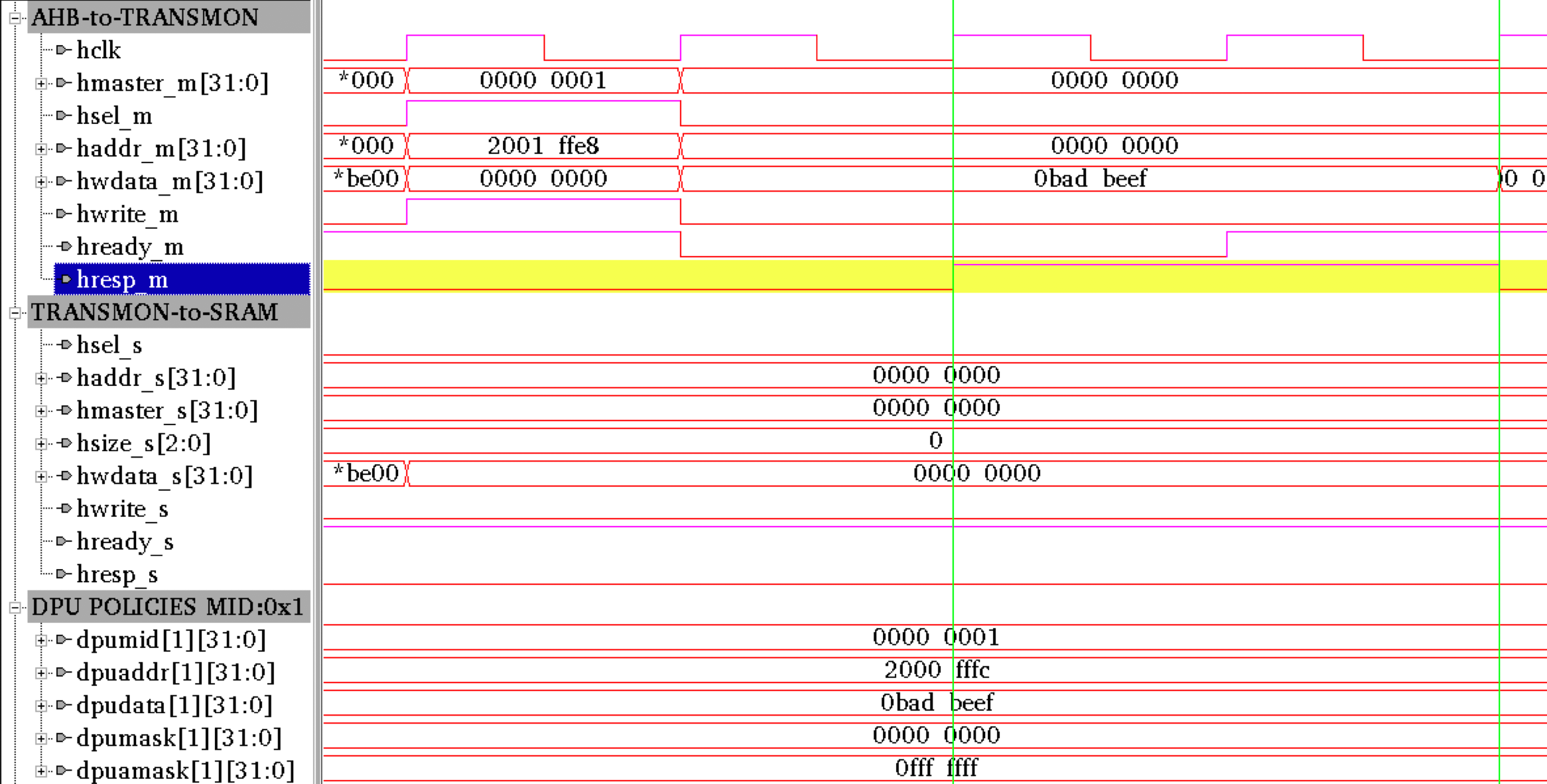}
}
\\
\subfloat[Malicious manipulation of data blocked by an APU policy.]{
	\includegraphics[height=6cm, trim = {0mm 0cm 8cm 0mm}, clip=true]{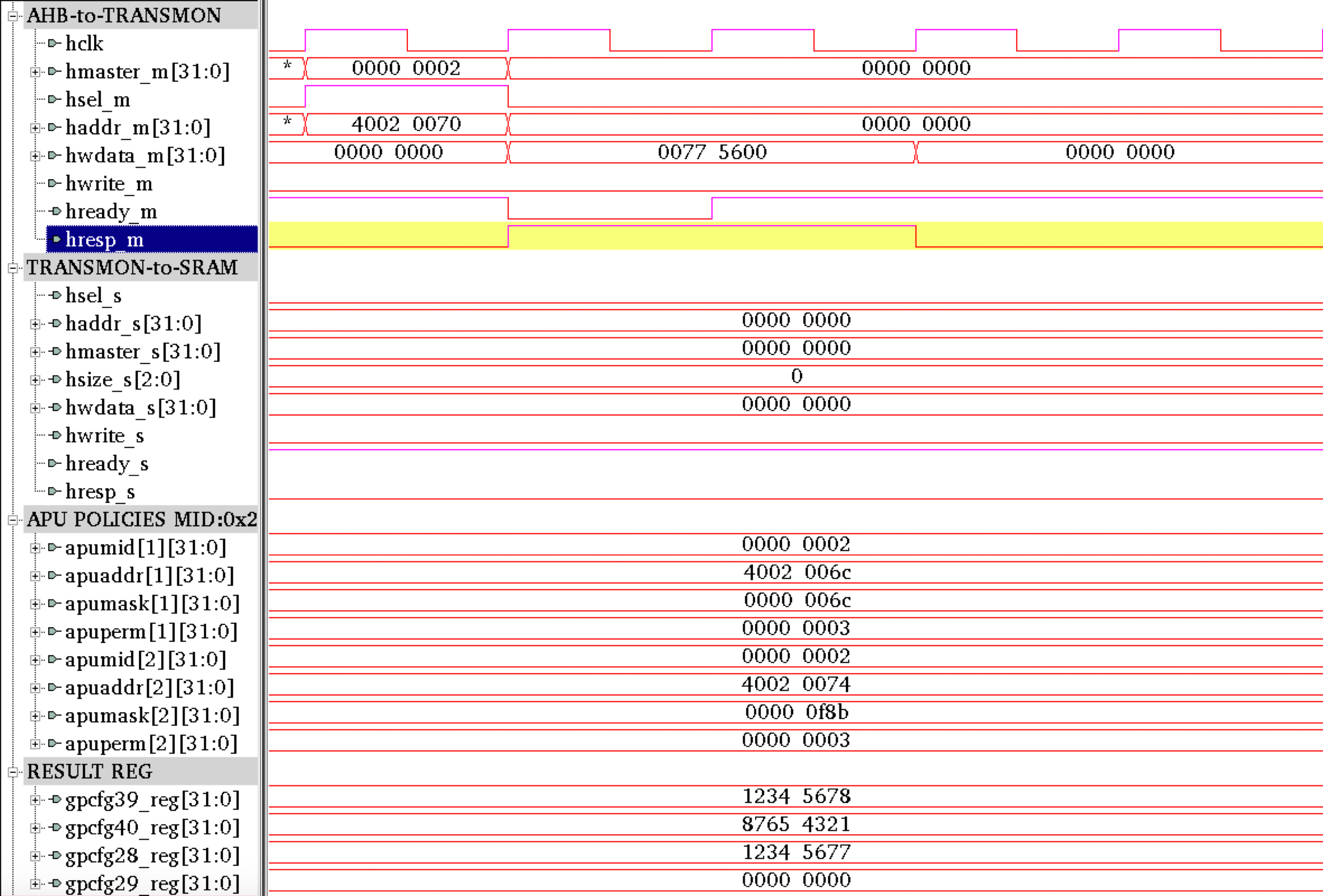}
}\hfill
\subfloat[Malicious over-writing of a semaphore in \textit{gpcfg39\_reg} blocked by a DPU policy.]{
	\includegraphics[height=5cm, trim = {0mm 0cm 0cm 0mm}, clip=true]{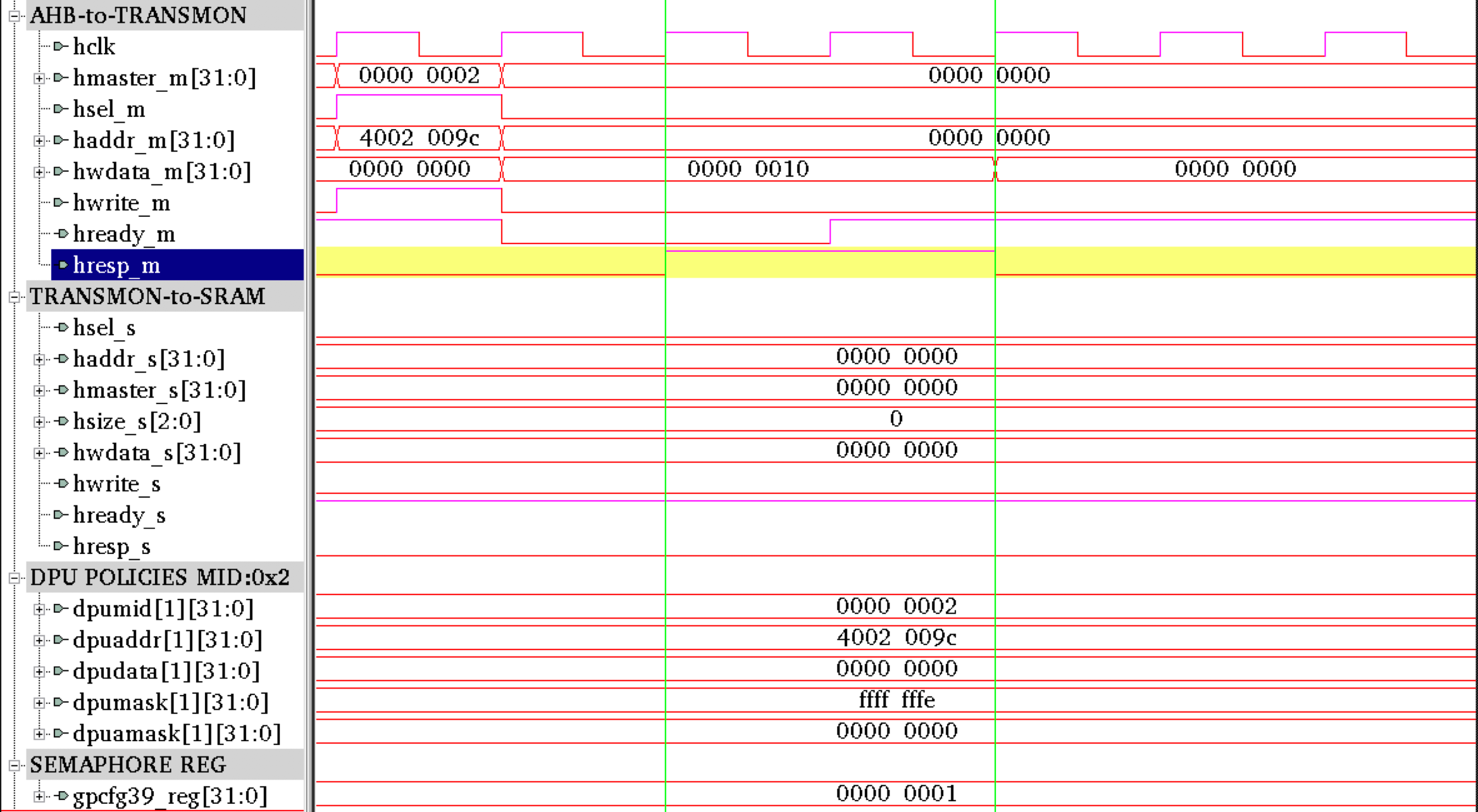}
}
\smallerspacecaption
\caption{Demonstration of ISEA at simulation level, obtained from \textit{Synopsys VCS}.
\label{fig:simulations}
}
\smallerspacecaption
\end{figure*}

Fig.~\ref{fig:simulations} illustrates the simulated behavior of ISEA.
We next outline the threat scenarios and related ISEA configurations, where
for all such scenarios ISEA successfully blocks all malicious transactions.

A malicious core tries to read/write some memory region previously declared as protected.
The related policy is concerning the address space; the APU covers this kind of threat. Waveform~\ref{fig:simulations}(a) shows how address protection policies help in blocking a transaction addressed to a protected region. An APU policy is set for the processor with ID 0x1, restricting its access by limiting it to the address range: 0x2000\_0000 (APUADDR AND NOT(APUMASK)) to 0x2000\_7FFF (APUADDR OR APUMASK). Any access outside this region will not be passed to the memory but, instead, an Error Response (\textit{hresp}) is returned to the master processor. Waveform~\ref{fig:simulations}(a) shows the Error Response generated for the address 0x2000\_F800 as it is outside the allowed range.

A malicious core tries to write out some \textit{sensitive asset}, e.g., a secret crypto key.
The related policy is concerning the actual data; the DPU covers this kind of threat. In Waveform~\ref{fig:simulations}(b), a DPU policy is set to track a write transaction by the master processor with ID 0x1 to the memory region between addresses 0x2000\_0000 (DPUADDR AND NOT(DPUAMASK)) to 0x2FFF\_FFFF (DPUADDR OR DPUAMASK) for the sensitive data 0x0BAD\_BEEF (DPUDATA AND NOT(DPUMASK)). The waveform also shows an attempt to write the restricted data value, which results in an Error Response generated by the DPU.

Two (or more) cores run the same computation in parallel, e.g., for cross-checking.
Then, one malicious core tries to (a) access the result of some other core, stored in the shared memory, and (b) possibly corrupt the other's result, to hide own malicious results from cross-checking.
The related policy is concerning the address space. Here the policy needs to be formulated such that access to the storage address of the other master is restricted.
Waveform~\ref{fig:simulations}(c) shows that the policy is formulated for the master ID 0x2 such that it has access to address ranges from 0x4002\_0000 (APUADDR[1] AND NOT(APUMASK[1])) to 0x4002\_006C (APUADDR[1] OR APUMASK[1]) and address ranges from 0x4002\_0074 (APUADDR[2] AND NOT(APUMASK[2])) to 0x4002\_0FFF (APUADDR[2] OR APUMASK[2]), but not to the address 0x4002\_0070, that is, where the master ID 0x1 stores its result. As covered in the waveform, with this setting of APU policies, when the master ID 0x2 tries to access the address 0x4002\_0070, the transaction is blocked by the APU, and an Error Response is returned. 

Two (or more) cores are establishing a \textit{semaphore} for software-based access control of shared memory regions.
Semaphores can be stored in the additional \textit{shared register space (SRS)} which is part of the ``root of trust,'' hence trustworthy by
itself.  Here, a malicious core tries to over-write the semaphore to be able to read/corrupt shared memory otherwise not accessible.
Hence, the related policy is concerning data access.
Waveform~\ref{fig:simulations}(d) shows how such a transaction is blocked. \textit{gpcfg39\_reg} is considered as a semaphore register here. For the processor master with ID 0x1, to obtain the ownership of semaphore, it has to write 0x0000\_0001 to the above register, but can do so only while the register value is 0x0000\_0000. For master ID 0x2, it similarly has to write 0x0000\_0010 to obtain the semaphore.
Naturally, one master should not be able to clear the bit set by any other master. A DPU policy is compiled to implement this data restriction. In the waveform, the DPU policy is set for the master ID 0x02 to restrict writing of ``0'' to the last bit (DPUDATA AND NOT(DPUMASK)) of the semaphore register. Then, the waveform shows an attempt to clear the last bit of the semaphore register by the master ID 0x2, resulting in the error response.

\subsection{Layout Evaluation}

We analyze and compare the PPA for three versions of the chip which include the non-secure version and two secure versions.
We implement the interposers at two different technology nodes, at 65nm and 90nm respectively. 
Table~\ref{tab:PPA} provides chiplet and interposer cell count and standard cell area contribution separately as C/I. %As expected, the interposer cell area for secure version is 1.3 times that of non-secure one. 
The power and delay overheads, when comparing the implementations (of secure and non-secure) at 65nm are 14.1\% and 20.2\% respectively.
The timing and power overheads between the interposer implementation at 65nm and 90nm is roughly 36\% and 13.6\% which is reflective of technology scaling~\cite{borkar99}.
Please note that area overhead in the interposer 
does not induce any additional silicon cost as the die area remains the same.

\begin{table}[tb]
\centering
\scriptsize
\caption{PPA results for different chip versions. C denotes chiplet and I denotes interposer.
}
\label{tab:PPA}
\smallerspacecaption
\smallerspacecaption
\setlength{\tabcolsep}{1mm}
\begin{tabular}{|c|c|c|c|c|}
\hline
 \textbf{Chip Version
 } & \textbf{Cell Count (C/I)} 
 & \textbf{Cell Area (C/I)} 
 & \textbf{Delay (ns)} 
 & \textbf{Power (mW)} \\ 
 \hline \hline
Non-secure  & 491,036/103,857 & 23,832,291/371,115 & 10.5 & 347 \\ \hline
Secure (Interposer 65nm) & 505,019/195,921  & 23,961,873/854,899 & 12.628  & 396 \\ \hline
Secure (Interposer 90nm) & 505,019/177,194 & 23,961,873/2,284,187  & 17.170  & 450 \\ \hline
\end{tabular}
\end{table}

\section{Conclusions and Future Work}
\label{sec:conclusion}

We propose and demonstrate a security concept providing clear physical separation at the system level. Our architecture, 
referred to as ISEA, 
is based on the 
2.5D technology, where, by leveraging 
an interposer as the secure entity, we enable trusted system-level integration of untrusted chiplets.
We propose different features within ISEA which protect against malicious data and address
transactions. 
As a proof-of-concept, we implement ISEA for a 64-core ARM M0 system and also explore physical design implementation for various foundry scenarios.
More specifically, we leverage 65nm and 90nm libraries by Globalfoundries and Synopsys.
Using an interposer is the only option to cost-effectively ``mix-and-match''
chiplets and technologies from different vendors and foundries---all while maintaining trust, as we have shown here.
We consider various threats and demonstrate related policies in action; ISEA can successfully act as ``police'' to enforce system-level security. 

As for future work, we note that other schemes like multi-party computation (MPC) inherently require a system-level security backbone. 
Hence, we shall also apply our scheme towards MPC. 
Currently,
our scheme protects against communication-level threats. However, we
envision the interposer as a general security platform, which can be augmented with various sensors as well. 
This way, we shall seek to identify malicious activities which are expressed by side-channels, but not detectable by security policies.



\begin{thebibliography}{10}
\providecommand{\url}[1]{#1}
\csname url@samestyle\endcsname
\providecommand{\newblock}{\relax}
\providecommand{\bibinfo}[2]{#2}
\providecommand{\BIBentrySTDinterwordspacing}{\spaceskip=0pt\relax}
\providecommand{\BIBentryALTinterwordstretchfactor}{4}
\providecommand{\BIBentryALTinterwordspacing}{\spaceskip=\fontdimen2\font plus
\BIBentryALTinterwordstretchfactor\fontdimen3\font minus
  \fontdimen4\font\relax}
\providecommand{\BIBforeignlanguage}[2]{{%
\expandafter\ifx\csname l@#1\endcsname\relax
\typeout{** WARNING: IEEEtran.bst: No hyphenation pattern has been}%
\typeout{** loaded for the language `#1'. Using the pattern for}%
\typeout{** the default language instead.}%
\else
\language=\csname l@#1\endcsname
\fi
#2}}
\providecommand{\BIBdecl}{\relax}
\BIBdecl

\bibitem{lipp18}
\BIBentryALTinterwordspacing
M.~Lipp \emph{et~al.}, ``Meltdown,'' \emph{ArXiv e-prints}, 2018. [Online].
  Available: \url{https://arxiv.org/abs/1801.01207}
\BIBentrySTDinterwordspacing

\bibitem{basak17}
A.~Basak \emph{et~al.}, ``Security assurance for system-on-chip designs with
  untrusted {IPs},'' \emph{Trans. Inf. Forens. Sec.}, vol.~12, no.~7, pp.
  1515--1528, 2017.

\bibitem{chandrasekharan15}
A.~Chandrasekharan \emph{et~al.}, ``Ensuring safety and reliability of
  {IP}-based system design -- a container approach,'' in \emph{Proc. Int. Symp.
  Rapid System Prototyping}, 2015, pp. 76--82.

\bibitem{wang15_IIPS}
X.~Wang \emph{et~al.}, ``{IIPS}: Infrastructure {IP} for secure {SoC} design,''
  \emph{Trans. Comp.}, vol.~64, no.~8, pp. 2226--2238, 2015.

\bibitem{liu15_TETC}
C.~Liu \emph{et~al.}, ``Shielding heterogeneous {MPSoCs} from untrustworthy
  {3PIPs} through security- driven task scheduling,'' \emph{Trans. Emerg. Top.
  Comp.}, vol.~2, no.~4, pp. 461--472, 2014.

\bibitem{fiorin08}
L.~Fiorin \emph{et~al.}, ``Secure memory accesses on networks-on-chip,''
  \emph{Trans. Comp.}, vol.~57, no.~9, pp. 1216--1229, 2008.

\bibitem{caimi17}
L.~L. Caimi \emph{et~al.}, ``Activation of secure zones in many-core systems
  with dynamic rerouting,'' in \emph{Proc. Int. Symp. Circ. Sys.}, 2017, pp.
  1--4.

\bibitem{wassel14}
H.~M.~G. Wassel \emph{et~al.}, ``Networks on chip with provable security
  properties,'' \emph{IEEE Micro}, vol.~34, no.~3, pp. 57--68, 2014.

\bibitem{rostami14}
M.~Rostami \emph{et~al.}, ``A primer on hardware security: Models, methods, and
  metrics,'' \emph{Proc. IEEE}, vol. 102, no.~8, pp. 1283--1295, 2014.

\bibitem{xiao16}
\BIBentryALTinterwordspacing
K.~Xiao \emph{et~al.}, ``Hardware trojans: Lessons learned after one decade of
  research,'' \emph{Trans. Des. Autom. Elec. Sys.}, vol.~22, no.~1, pp.
  6:1--6:23, 2016. [Online]. Available:
  \url{http://doi.acm.org/10.1145/2906147}
\BIBentrySTDinterwordspacing

\bibitem{subramanyan14}
P.~Subramanyan \emph{et~al.}, ``Reverse engineering digital circuits using
  structural and functional analyses,'' \emph{Trans. Emerg. Top. Comp.},
  vol.~2, no.~1, pp. 63--80, 2014.

\bibitem{wang17_data}
L.~C. Wang, ``Experience of data analytics in {EDA} and test - principles,
  promises, and challenges,'' \emph{Trans. Comp.-Aided Des. Integ. Circ. Sys.},
  vol.~PP, no.~99, pp. 1--1, 2017.

\bibitem{valamehr10}
J.~Valamehr \emph{et~al.}, ``Hardware assistance for trustworthy systems
  through {3-D} integration,'' in \emph{Proc. Ann. Comp. Sec. App. Conf.},
  2010, pp. 199--210.

\bibitem{valamehr13}
------, ``A {3-D} split manufacturing approach to trustworthy system
  development,'' \emph{Trans. Comp.-Aided Des. Integ. Circ. Sys.}, vol.~32,
  no.~4, pp. 611--615, 2013.

\bibitem{clermidy16}
F.~Clermidy \emph{et~al.}, ``New perspectives for multicore architectures using
  advanced technologies,'' in \emph{Proc. Int. Elec. Devices Meeting}, 2016,
  pp. 35.1.1--35.1.4.

\bibitem{pavlidis17}
V.~F. Pavlidis \emph{et~al.}, \emph{Three-dimensional Integrated Circuit
  Design}, 2nd~ed.\hskip 1em plus 0.5em minus 0.4em\relax Morgan Kaufmann
  Publishers Inc., 2017.

\bibitem{lau11}
J.~H. Lau, ``The most cost-effective integrator ({TSV} interposer) for {3D IC}
  integration system-in-package ({SiP}),'' in \emph{Proc. ASME InterPACK},
  2011, pp. 53--63.

\bibitem{stow17}
D.~Stow \emph{et~al.}, ``Cost-effective design of scalable high-performance
  systems using active and passive interposers,'' in \emph{Proc. Int. Conf.
  Comp.-Aided Des.}, 2017.

\bibitem{lee16}
C.~C. Lee \emph{et~al.}, ``An overview of the development of a {GPU} with
  integrated {HBM} on silicon interposer,'' in \emph{Proc. Elec. Compon. Tech.
  Conf.}, 2016, pp. 1439--1444.

\bibitem{shilov18}
\BIBentryALTinterwordspacing
A.~Shilov. (2018) {AMD} previews {EPYC} rome processor: Up to 64 {Zen 2} cores.
  [Online]. Available:
  \url{https://www.anandtech.com/show/13561/amd-previews-epyc-rome-processor-up-to-64-zen-2-cores}
\BIBentrySTDinterwordspacing

\bibitem{CHIPS}
\BIBentryALTinterwordspacing
(2016) Common heterogeneous integration and intellectual property ({IP}) reuse
  strategies ({CHIPS}). [Online]. Available:
  \url{https://www.darpa.mil/program/common-heterogeneous-integration-and-ip-reuse-strategies}
\BIBentrySTDinterwordspacing

\bibitem{yin18}
J.~Yin \emph{et~al.}, ``Modular routing design for chiplet-based systems,'' in
  \emph{Proc. Int. Symp. Comp. Archit.}, 2018, pp. 726--738.

\bibitem{coskun18}
A.~Coskun \emph{et~al.}, ``A cross-layer methodology for design and
  optimization of networks in {2.5D} systems,'' in \emph{Proc. Int. Conf.
  Comp.-Aided Des.}, 2018.

\bibitem{mahajan16}
R.~Mahajan \emph{et~al.}, ``Embedded multi-die interconnect bridge ({EMIB}) --
  a high density, high bandwidth packaging interconnect,'' in \emph{Proc. Elec.
  Compon. Tech. Conf.}, 2016, pp. 557--565.

\bibitem{takaya13}
S.~Takaya \emph{et~al.}, ``A 100gb/s wide i/o with 4096b tsvs through an active
  silicon interposer with in-place waveform capturing,'' in \emph{Proc. Int.
  Sol.-St. Circ. Conf.}, 2013, pp. 434--435.

\bibitem{akgun16}
I.~Akgun \emph{et~al.}, ``Scalable memory fabric for silicon interposer-based
  multi-core systems,'' in \emph{Proc. Int. Conf. Comp. Des.}, 2016, pp.
  33--40.

\bibitem{kannan15}
A.~Kannan \emph{et~al.}, ``Enabling interposer-based disintegration of
  multi-core processors,'' in \emph{Proc. Int. Symp. Microarch.}, 2015, pp.
  546--558.

\bibitem{indrusiak16}
\BIBentryALTinterwordspacing
L.~S. Indrusiak \emph{et~al.} (2016, July) Side-channel attack resilience
  through route randomisation in secure real-time networks-on-chip. arXiv.
  [Online]. Available: \url{http://arxiv.org/abs/1607.03450}
\BIBentrySTDinterwordspacing

\bibitem{borkar99}
S.~Borkar, ``Design challenges of technology scaling,'' \emph{IEEE Micro},
  vol.~19, no.~4, pp. 23--29, 1999.

\end{thebibliography}
\end{document}